# Additional Waves and Additional Boundary Conditions in Local Quartic Metamaterials

Morgan LaBalle[1], Maxim Durach[1*]

1. The Department of Physics and Astronomy, Georgia Southern University

*email: mdurach@georgiasouthern.edu

**Abstract:** Additional electromagnetic waves and additional boundary conditions (ABCs) in non-local materials attracted a lot of attention in the past. Here we report the possibility of additional propagating and evanescent waves in *local* anisotropic and bi-anisotropic linear materials. We investigate the possible options for ABCs and describe how to complement the conventional 4 Maxwell's boundary conditions in the situations when there are more than 4 waves that need to be matched at the boundary of local and linear quartic metamaterials. We show that these ABCs must depend on the properties of the interface and require the introduction of the additional effective material parameters describing this interface, such as surface conductivities.

## I. Introduction

Optical fields in conventional local optical materials follow quadratic dispersion conditions, e.g. $k^2 = k_0^2$ in vacuum. The k-vectors of the plane waves that compose these fields belong to quadratic surfaces – spheres, hyperboloids, ellipsoids, i.e. such k-surfaces (or iso-frequency surfaces) are described by the second order polynomials in k-vector components for any frequency [1,2]. In such materials no more than 2 plane waves are allowed to propagate or decay in any direction, and at boundaries between such media the usual Maxwell's boundary conditions (MBCs) form complete systems of equations for the amplitudes of the fields. As was first pointed out by Pekar [3,4] in the presence of spatial dispersion it is possible to excite more than 2 plane waves propagating in the same direction, and the appearance of the additional waves calls for additional boundary conditions (ABCs) at the boundaries of media with spatial dispersion. So far the investigations and the use of the ABCs has been limited to media with spatial dispersion, such as semiconductors with excitons [5] or wire metamaterials [1,2,6-9]. Independently, the development of the metamaterial cloaking, the field of optical metasurfaces and optics of 2D materials has presented the need for the modification of the MBCs themselves by including the discontinuity of the tangential components of the magnetic field **H** at the cloaking mantle [10], metasurface [11,12] or graphene sheet [13,14], which is associated with their finite optical surface conductivity.

In this paper we first demonstrate that the additional waves may theoretically arise in *local* anisotropic or bi-anisotropic media and then discuss the corresponding ABCs for the boundaries of these media. We show that these ABCs must depend on the properties of the interface and require the introduction of the additional effective material parameters describing this interface, such as surface charges and conductivities, which correspond to the discontinuities of the normal components of fields **E, H** and tangential components of the fields **D, B**. We demonstrate that such interfaces can support electromagnetic surface waves that are composed from up to 8 evanescent waves in the interfacing media.

Here we focus on bi-anisotropic quartic metamaterials with material relation of the form [15-20]

$$\begin{pmatrix}D\\B\end{pmatrix} = \widehat{M}\begin{pmatrix}E\\H\end{pmatrix}, \widehat{M} = \begin{pmatrix}\hat{\epsilon} & \hat{X}\\ \hat{Y} & \hat{\mu}\end{pmatrix}, \quad (1)$$

where $\hat{\epsilon}$, $\hat{\mu}$, $\hat{X}$, $\hat{Y}$ are 3x3 tensors characterizing dielectric permittivity, magnetic permeability and magnetoelectric coupling. The dispersion relation in such materials is given by the quartic equation [21]

$$f(k_x, k_y, k_z) = \sum_{i+j+l+m=4}[\alpha_{ijlm} k_x^i k_y^j k_z^l k_0^m] = 0 \quad (2)$$

with 35 coefficients $\alpha_{ijlm}$. In the sum the powers $i$, $j$, $l$, $m$ run from 0 to 4 such that $i + j + l + m = 4$.

## II. Additional Waves in Local Metamaterials

We would like to start by demonstrating that even anisotropic local materials without magnetoelectric coupling can feature additional waves. Indeed, if for some direction in such a material, say for y-direction, the projection of $\boldsymbol{D}$ and/or $\boldsymbol{B}$ vectors are identically zero $D_y = 0, B_y = 0$ for any electric and magnetic fields $\boldsymbol{E}, \boldsymbol{H}$, then this should be expressed as $\epsilon_{21} = \epsilon_{22} = \epsilon_{23} = 0$ and/or $\mu_{21} = \mu_{22} = \mu_{23} = 0$. In this circumstance any k-vector in this direction $\boldsymbol{k} = (0, k_y, 0)$ gives a solution of Maxwell's equation, since the corresponding system of equations reduces to 4 or 5 equations for 6 components of vectors $\boldsymbol{E}, \boldsymbol{H}$.

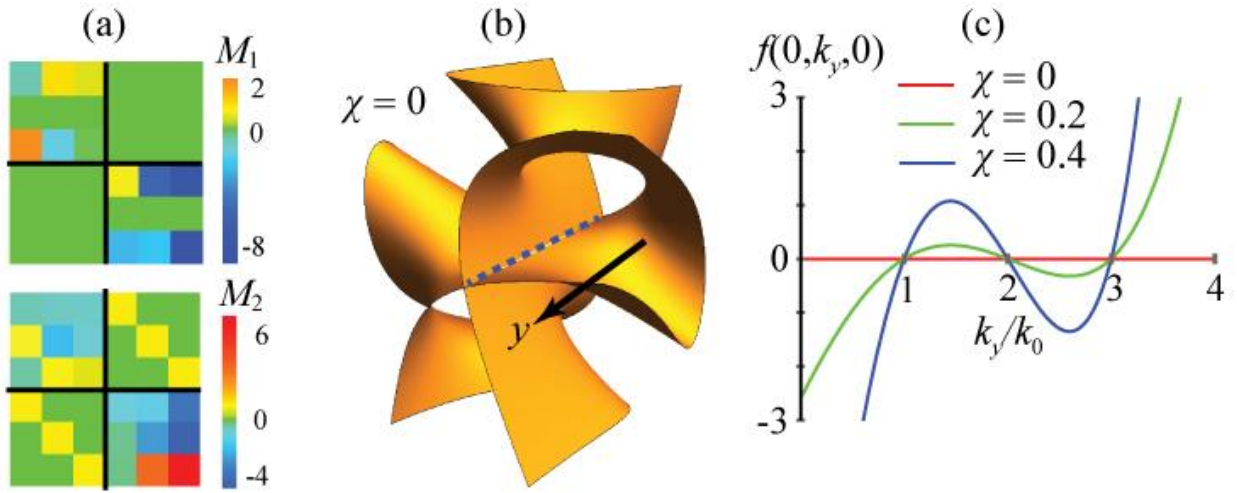

Fig. 1. Additional waves in a local anisotropic material. (a) effective parameters matrix $\widehat{M} = \widehat{M}_1 + \chi\widehat{M}_2$; (b) The k-surface for $\chi = 0$ with infinite number of additional waves [dashed blue line]; (c) reduction of the number of additional waves with introduction of the magnetoelectric coupling: 3 waves propagate in the positive y-direction, i.e. 1 additional wave.

For example, consider a material with material relation $\hat{M} = \hat{M}_1 + \chi\hat{M}_2$, where matrices $\hat{M}_1$ and $\hat{M}_2$ are given in Fig. 1(a). For $\chi = 0$ this material has an infinite continuum of additional waves with $\boldsymbol{k} = (0, k_y, 0)$ [dashed blue line in Fig. 1(b)]. For non-zero $\chi$, which introduces magnetoelectric coupling, the number of waves propagating in the positive y-direction reduces to 3, i.e. there is one additional wave in such materials [Fig. 1(c)]. Generally, one can have up to 2 additional waves in local quartic metamaterials as shown in Fig. 2. In Fig. 2(a) we show an example of k-surface that has 4 waves (shown as red dots) with phase propagating in the same direction (blue line), which corresponds to 2 additional waves, and the corresponding effective parameter matrix $\hat{M}$.

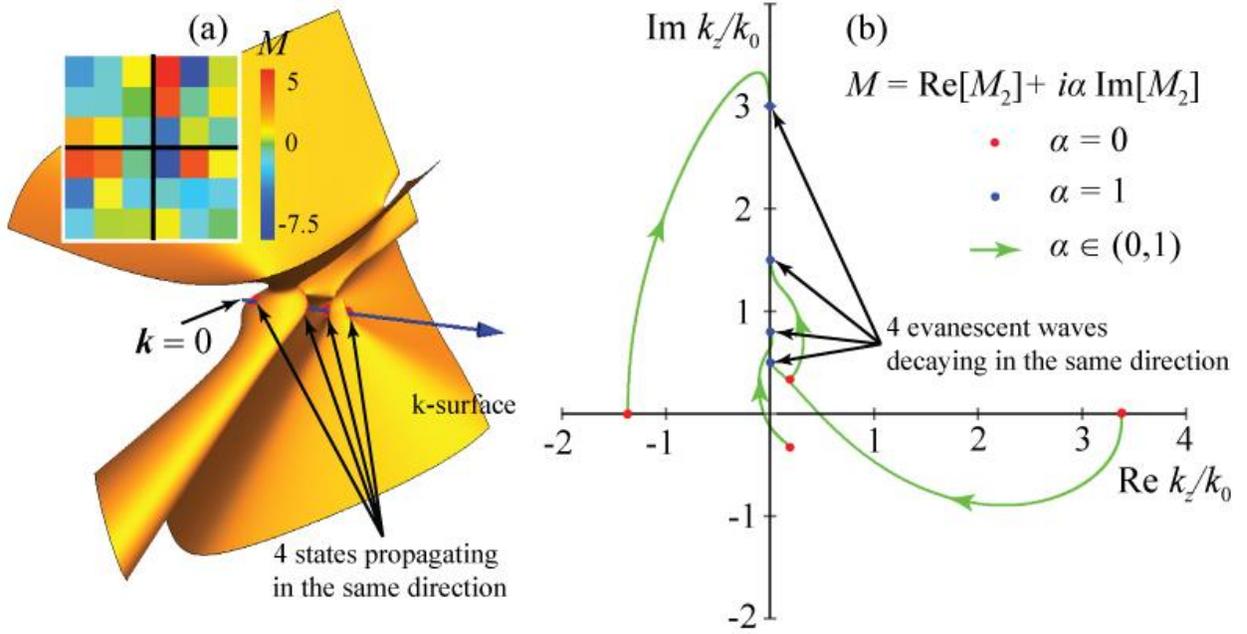

Fig. 2. Propagating and evanescent additional waves. (a) The k-surface for the material shown in the inset, which propagates the phases of 4 waves (red dots) in the same direction (blue line). (b) Complex $k_z$-plane which shows the evolution of the waves with k = $(1.5k_0, 0, k_z)$, when the effective material parameters matrix $\hat{M} = \text{Re}[\hat{M}_2] + i\alpha\,\text{Im}[\hat{M}_2]$ is changed from $\alpha = 0$ (red dots) to $\alpha = 1$ (blue dots), where matrix $\hat{M}_2$ comes from Fig. 4(a).

The additional evanescent waves require a slightly different consideration. Consider evanescent waves that decay in z direction with $\boldsymbol{k} = (\boldsymbol{k}_\parallel, k_z)$. If we fix $\boldsymbol{k}_\parallel = (k_x, k_y)$, then Eq.(2) becomes a single-variable quartic equation with respect to a complex $k_z = k_z' + ik_z''$. If the coefficients $\alpha_{ijlm}$ are real then the quartic equation for $k_z$ has 4 roots which are real or come in complex conjugated pairs, i.e. for each decaying wave in the positive z-direction there is a growing wave, without additional waves. From this we conclude that for additional evanescent waves we need complex coefficients $\alpha_{ijlm}$ which cannot be achieved, for example, in materials with real material parameters $\hat{M}$. In Fig. 2(b) we show how a material with real matrix $\hat{M}_r$ acquires 2 additional evanescent waves if an imaginary part is added to the material parameters matrix $\hat{M} = \hat{M}_r + i\alpha\hat{M}_i$. For the numerical example we use matrix $\hat{M}_2$ from Fig. 4(a) such that $\hat{M}_r =$

Re$[\widehat{M}_2]$ and $\widehat{M}_i = \text{Im}[\widehat{M}_2]$ and we change $\alpha$ from 0 to 1. For $\alpha = 0$ (red dots) there are 2 real roots, corresponding to waves propagating phase in the positive and negative z-directions, and 2 roots that are complex conjugated of each other, i.e. the related waves decay in the positive and negative z-directions, with no additional waves. For $\alpha = 1$ (blue dots) there are 4 waves all decaying in the positive z-direction, i.e. 2 additional evanescent waves.

### III.    Additional Boundary Conditions for Local Metamaterials

Now let us turn to a boundary between two quartic metamaterials $M_1$ ($z < 0$) and $M_2$ ($z > 0$) located in *xy*-plane at $z = 0$. Solutions of Maxwell's equations with different longitudinal k-vectors $\boldsymbol{k}_\parallel = (k_x, k_y)$ in each material are linearly independent due to translational symmetry. For a fixed $\boldsymbol{k}_\parallel$ Eq.(2) is a quartic equation of the single variable $k_z$. The 4 roots of this equation are complex $\boldsymbol{k} = (\boldsymbol{k}_\parallel, k'_z + ik''_z)$ and represent propagating or evanescent waves. Since $k_z$ can be chosen arbitrarily [21], we first focus on the extreme case when all 4 of them are with $\text{Im } k_z \leq 0$ in $M_1$, and all 4 with $\text{Im } k_z \geq 0$ in $M_2$, so that all 8 waves can be a part of the field at the boundary and we have 4 additional waves in total. We proceed by selecting the custom field components for these waves as $\Gamma_i^{(1,2)} = \left(\Gamma_{\parallel i}^{(1,2)}, \Gamma_{\perp i}^{(1,2)}\right)$, where $\Gamma_{\parallel i}^{(j)} = \left(E_{xi}^{(j)}, E_{yi}^{(j)}, H_{xi}^{(j)}, H_{yi}^{(j)}\right)$ are longitudinal fields and $\Gamma_{\perp i}^{(j)} = \left(E_{zi}^{(j)}, H_{zi}^{(j)}\right)$ are transverse fields. The MBCs require the continuity of the longitudinal components of the fields, which results in the following 4 equations:

$$A\Gamma_{\parallel 1}^{(1)} + B\Gamma_{\parallel 2}^{(1)} + C\Gamma_{\parallel 3}^{(1)} + D\Gamma_{\parallel 4}^{(1)} = E\Gamma_{\parallel 1}^{(2)} + F\Gamma_{\parallel 2}^{(2)} + G\Gamma_{\parallel 3}^{(2)} + H\Gamma_{\parallel 4}^{(2)}, \qquad (3)$$

for 8 unknown coefficients $A, B, C, D, E, F, G, H$ which are the amplitudes of the waves composing the field at the quartic boundary.

We start by considering a boundary between two quartic media with matrices $\widehat{M}$ shown in Fig. 3(a). For $\boldsymbol{k}_\parallel = (k_0, 0)$ these materials support waves with $k_z = k_0; -\frac{k_0}{2}; -\frac{ik_0}{2}; -6ik_0$ in $M_1$ and $k_z = \frac{k_0}{2}; 3k_0; ik_0; 5ik_0$ in $M_2$. Only one of the waves propagates toward the boundary - $\Gamma_1^{(1)}$, with one possible reflection $\Gamma_2^{(1)}$, two possible transmissions $\Gamma_{1,2}^{(2)}$ and 4 evanescent waves decaying from the boundary $\Gamma_{3,4}^{(1,2)}$. Setting $A = 1$ fixes the incident power, which together with Eq.(3) leaves 3 amplitudes of the waves undefined. This means that we can select any 3 amplitudes of the additional waves arbitrarily and obtain a continuum of solutions all satisfying Maxwell's equations. We demonstrate this in Fig. 3(b-d), where we show 3 possible field distributions for these materials. In Fig. 3(b) we show the excitation of the new kind of surface plasmon polaritons that has been recently proposed in Ref. [22] by setting the amplitudes of reflected and transmitted waves to zero $B = E = F = 0$. We then show the possibility of the selection between the two transmitted waves by setting the amplitudes of one of them to zero in Figs. 3(c-d).

The multitude of the possible solutions of Maxwell's equations which arise at the boundary of two identical media upon identical excitation means that the difference between these solutions stems from the difference at the interface between the media on the microscopic level, which evades the description by the bulk effective medium approximation. Below we discuss the ways of complementing this bulk effective medium description by the additional surface effective medium description using the ABCs.

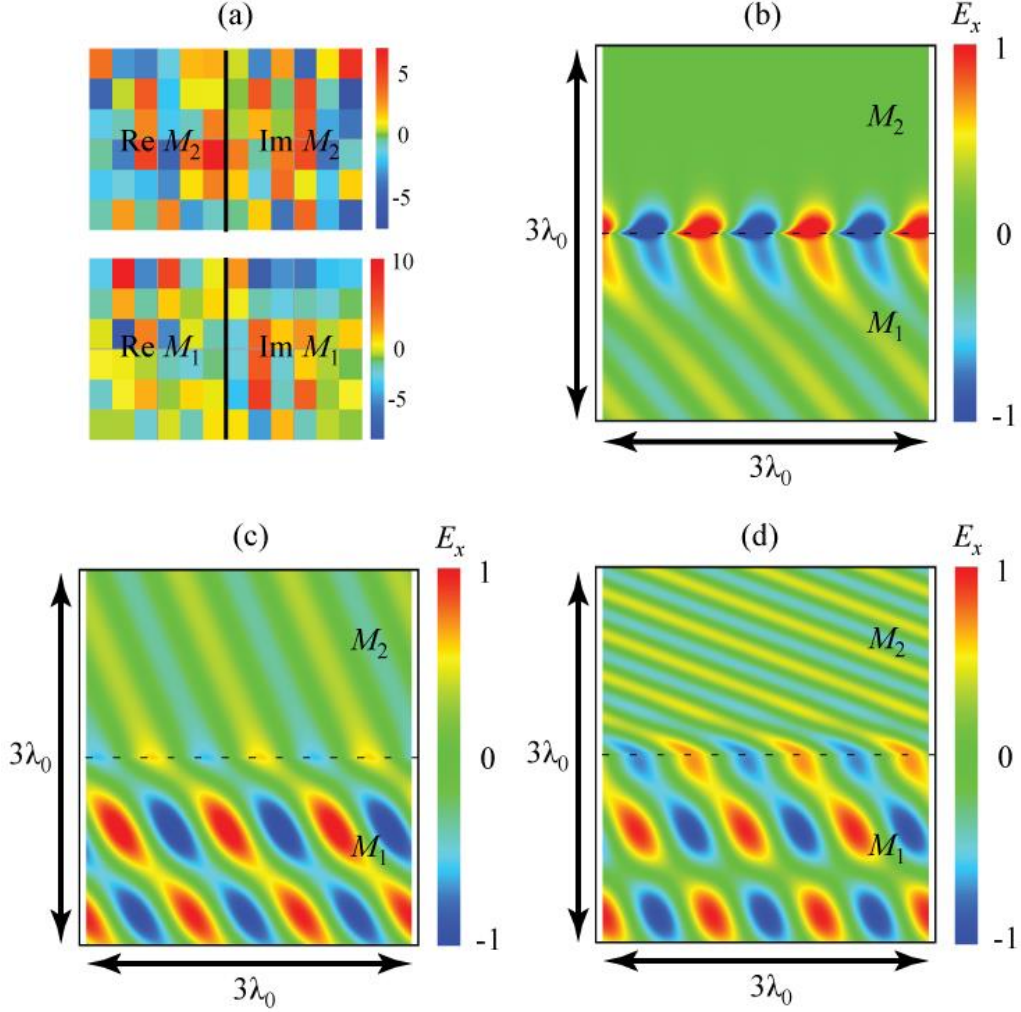

Fig. 3. Possible solutions of Maxwell's equations at the boundary of quartic metamaterials, whose effective parameters matrices are shown in (a); (b)-(d) illumination of the interfaces between identical quartic metamaterials with the identical incident wave resulting in excitation of (a) a new kind of surface plasmon polaritons (see Ref. [22]) and (c-d) two different transmitted waves depending on selection of the amplitudes of the additional waves.

Eqs. (3) are written for the continuous components of the fields. The discontinuities of the other components of the fields can be ascribed to the electric and magnetic surface charges $\sigma_e, \sigma_m$ and surface currents $\boldsymbol{j}_{\|e}, \boldsymbol{j}_{\|m}$

$$\boldsymbol{B}_\|^{(2)} - \boldsymbol{B}_\|^{(1)} = \frac{4\pi}{c}\boldsymbol{j}_{\|e}, \quad (4a); \qquad \boldsymbol{D}_\|^{(2)} - \boldsymbol{D}_\|^{(1)} = \frac{4\pi}{c}\boldsymbol{j}_{\|m}, \quad (4b)$$

$$E_z^{(2)} - E_z^{(1)} = 4\pi\sigma_e, \quad (5a); \qquad H_z^{(2)} - H_z^{(1)} = 4\pi\sigma_m, \quad (5b)$$

These surface charges and currents can be considered as induced by the corresponding continuous fields and should be proportional to them in the linear approximation

$$\begin{pmatrix} j_{\|e} \\ j_{\|m} \end{pmatrix} = \hat{\Sigma} \begin{pmatrix} E_\| \\ H_\| \end{pmatrix} = \begin{pmatrix} \hat{\sigma}_e & \hat{\xi} \\ \hat{\zeta} & \hat{\sigma}_m \end{pmatrix} \begin{pmatrix} E_\| \\ H_\| \end{pmatrix}, \quad (6a); \quad \begin{pmatrix} \sigma_e \\ \sigma_m \end{pmatrix} = \hat{\Lambda} \begin{pmatrix} D_z \\ B_z \end{pmatrix} = \begin{pmatrix} \alpha & \beta \\ \gamma & \delta \end{pmatrix} \begin{pmatrix} D_z \\ B_z \end{pmatrix}. \quad (6b)$$

The matrices $\hat{\Sigma}$ and $\hat{\Lambda}$ provide the effective description of the interface between quartic materials and specify the corresponding solutions for the distributions of the fields. The 4x4 matrix $\hat{\Sigma}$ contains electric and magnetic surface conductivity matrices $\hat{\sigma}_e$ and $\hat{\sigma}_m$ and the surface magnetoelectric couplings $\hat{\xi}$ and $\hat{\zeta}$. The 2x2 matrix $\hat{\Lambda}$ relates the discontinuities in the normal components $E_z$ and $H_z$ to the normal components of continuous $D_z$ and $B_z$.

Using all of the Eqs. (4-5) makes the system of equations for field amplitudes overdetermined, since there can be at most 4 additional waves in quartic metamaterials. Nevertheless Eqs. (4-5) provide the options for the selection of ABCs according to the problem at hand. For example, in the case of the materials and fields shown in Fig. 3 there are 3 additional waves and only 3 ABCs are needed. Therefore, we could select Eqs. (4a) and (5a) as ABCs. Then the effective surface description of Eq. (6) should be reduced to the case of $\hat{\sigma}_m = \hat{\xi} = \hat{\zeta} = 0$ and $\beta = \gamma = \delta = 0$.

Below we select the surface conductivity $\hat{\sigma}_e$ to be diagonal and find its components as well as $\alpha$ for the field distributions in Fig. 3(b-d) as demonstrated in Table 1.

| Table 1 | Fig.3(b) | Fig.3(c) | Fig.3(d) |
|---|---|---|---|
| $\Delta B_\|$ | $(17.91 + 24.03i, 32.08 + 0.36i)$ | $(0.26 - 0.73i, 1.62 + 0.72i)$ | $(-1.33 - 3.66i, 0.73 - 0.90i)$ |
| $E_\|$ | $(-2.78 - 3.49i, -3.38 + 1.22i)$ | $(0.58 - 0.44i, 1.05 + 0.12i)$ | $(-0.46 - 0.92i, 0.74 + 0.64i)$ |
| $\frac{4\pi}{c}\hat{\sigma}_e$ | $\begin{pmatrix} -6.70 - 0.23i & 0 \\ 0 & -8.37 - 3.14i \end{pmatrix}$ | $\begin{pmatrix} 0.90 - 0.57i & 0 \\ 0 & 1.60 + 0.50i \end{pmatrix}$ | $\begin{pmatrix} 3.76 + 0.43i & 0 \\ 0 & -0.036 - 1.18i \end{pmatrix}$ |
| $\Delta E_z$ | $-1.54 - 3.77i$ | $-1.63 + 0.29i$ | $-2.02 - 0.066i$ |
| $D_z$ | $4.68 - 0.81i$ | $0.09 - 0.25i$ | $1.18 + 1.41i$ |
| $4\pi\alpha$ | $-0.19 - 0.84i$ | $-3.05 - 5.33i$ | $-0.73 + 0.82i$ |

This illustrates that the different fields that arise from the identical excitation at the boundary of identical quartic metamaterials can be attributed to the difference of the interface, expressed via surface material parameters matrices $\hat{\Sigma}$ and $\hat{\Lambda}$.

Consider the extreme circumstance when 4 additional waves are present at a boundary of quartic metamaterials. In this case the excitation of the fields cannot come from the bulk of the media and therefore this case corresponds to an extreme example of a surface plasmon polariton composed of 8 evanescent waves. In this situation we choose to complement the MBCs of Eq. (1) with the ABCs of Eq. (4) and characterize the interface with the matrix $\hat{\Sigma}$ (Eqs. (4) and (6)). The condition of the existence of the field at such a boundary corresponds to vanishing of the determinant

$$\begin{vmatrix} \hat{G}_\|^{(1)} & -\hat{G}_\|^{(2)} \\ \hat{P}_\|^{(1)} - \frac{4\pi}{c}\hat{\Sigma}\hat{G}_\|^{(1)} & -\hat{P}_\|^{(2)} \end{vmatrix} = 0, \tag{7}$$

where matrix $\hat{G}_\|^{(j)}$ is composed of 4 vectors $\Gamma_{\|i}^{(j)}, i = 1-4$ in each medium, while $P_\|^{(j)}$ is composed of vectors $\Pi_{\|i}^{(j)} = \left(B_{xi}^{(j)}, B_{yi}^{(j)}, D_{xi}^{(j)}, D_{yi}^{(j)}\right)$ in a similar fashion.

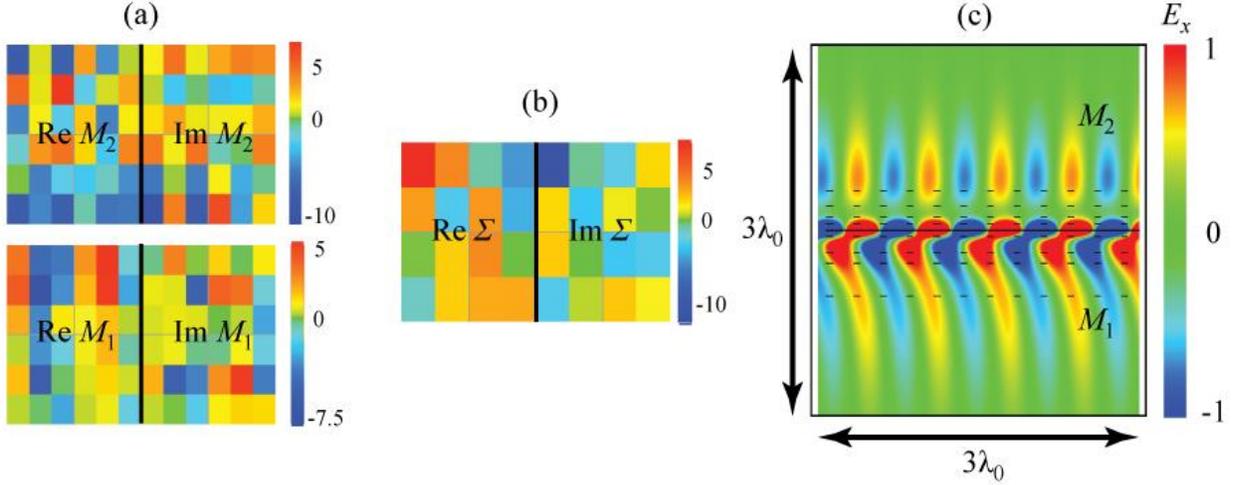

Fig. 4. The surface plasmon polariton composed of 8 evanescent waves, i.e. containing 4 additional evanescent waves. (a) The effective material parameters matrix $\hat{M}$. (b) The effective surface material parameters matrix $\hat{\Sigma}$. (c) The longitudinal electric field of the surface plasmon polariton, illustrating the interference of 4 evanescent waves in each interfacing material. The dashed lines show the decay lengths of the evanescent waves measured from the boundary (solid black line)

In Fig. 4 we demonstrate the formation of such SPP mode at the boundary of 2 randomly selected quartic metamaterials $M_1$ ($z < 0$) and $M_2$ ($z > 0$) with an interface described by randomly selected $\hat{\Sigma}$, such that Eq. (7) is fulfilled.

In conclusion, we show that local optical materials support additional propagating and evanescent waves, such that identical excitation of a boundary of identical quartic materials permits a continuum of different response fields in accordance to Maxwell's boundary conditions. We introduce additional boundary conditions that are based on the surface effective material parameters, which allows differentiating the response fields based on the surface currents and surface charges induced at the boundary.

**References.**